\documentclass{sig-alternate}
\conferenceinfo{}{}
\CopyrightYear{2016} 
\setcopyright{acmcopyright}
\conferenceinfo{SIGIR '16,}{July 17-21, 2016, Pisa, Italy}
\isbn{978-1-4503-4069-4/16/07}
\acmPrice{\$15.00}
\doi{http://dx.doi.org/10.1145/2911451.2911465}

\begin{document}
%

\title{Expedition: A Time-Aware Exploratory Search System Designed for Scholars}
%
%
%
%
%

\numberofauthors{1}
 \author{
 \alignauthor
 Jaspreet Singh, Wolfgang Nejdl, Avishek Anand\\
        \affaddr{L3S Research Center, Leibniz Universit\"at Hannover}\\
        \affaddr{Appelstr. 9a}\\
        \affaddr{30167 Hanover, Germany}\\
        \email{\{singh,nejdl,anand\}@L3S.de}
 }

\maketitle



\maketitle
\begin{abstract}


Archives are an important source of study for various scholars. Digitization and the web have made archives more accessible and led to the development of several time-aware exploratory search systems. However these systems have been designed for more general users rather than scholars. Scholars have more complex information needs in comparison to general users. They also require support for corpus creation during their exploration process. In this paper we present \textsf{Expedition} - a time-aware exploratory search system that addresses the requirements and information needs of scholars. \textsf{Expedition} possesses a suite of ad-hoc and diversity based retrieval models to address complex information needs; a newspaper-style user interface to allow for larger textual previews and comparisons; entity filters to more naturally refine a result list and an interactive annotated timeline which can be used to better identify periods of importance.

\end{abstract}




\section{Introduction}





Archives are an invaluable source of study for a variety of scholars like historians, social scientists, political scientists and journalists. The proliferation of such longitudinal corpora has seen significant strides made in Temporal IR  - new retrieval models, temporal profiling techniques and various indexing methods have been proposed to address the challenges posed when searching such collections~\cite{Campos:2014:STI:2658850.2619088}. This in turn has led to the development of several specialized search and exploration systems where time is a first class citizen. Time-aware exploratory search systems primarily utilize either (i) temporal retrieval models to improve search result ranking~\cite{mishra_expose:_2015} or (ii) timelines to detect trends and visually explore the collection~\cite{alonso_clustering_2009,setty2010inzeit,Alonso:2010:NNE:2128344.2128429}. However these systems are designed for more general users rather than the primary user group of digital archives - scholars. 

Scholars differ from more general users in the following ways: (i) a scholar's information need tends to be more complex. A complex information need requires the user to issue several queries whose intent varies even during the exploration process. For instance, initially the user may want an overview and subsequently more focused results. (ii) While exploring the archive, scholars track relevant results to form a corpus of study (iii) scholars further analyze this corpus and then publish not only their results but also the corpus construction procedure. 

\begin{figure*}[t]  
  \centering 
    \includegraphics[width=0.91\textwidth]{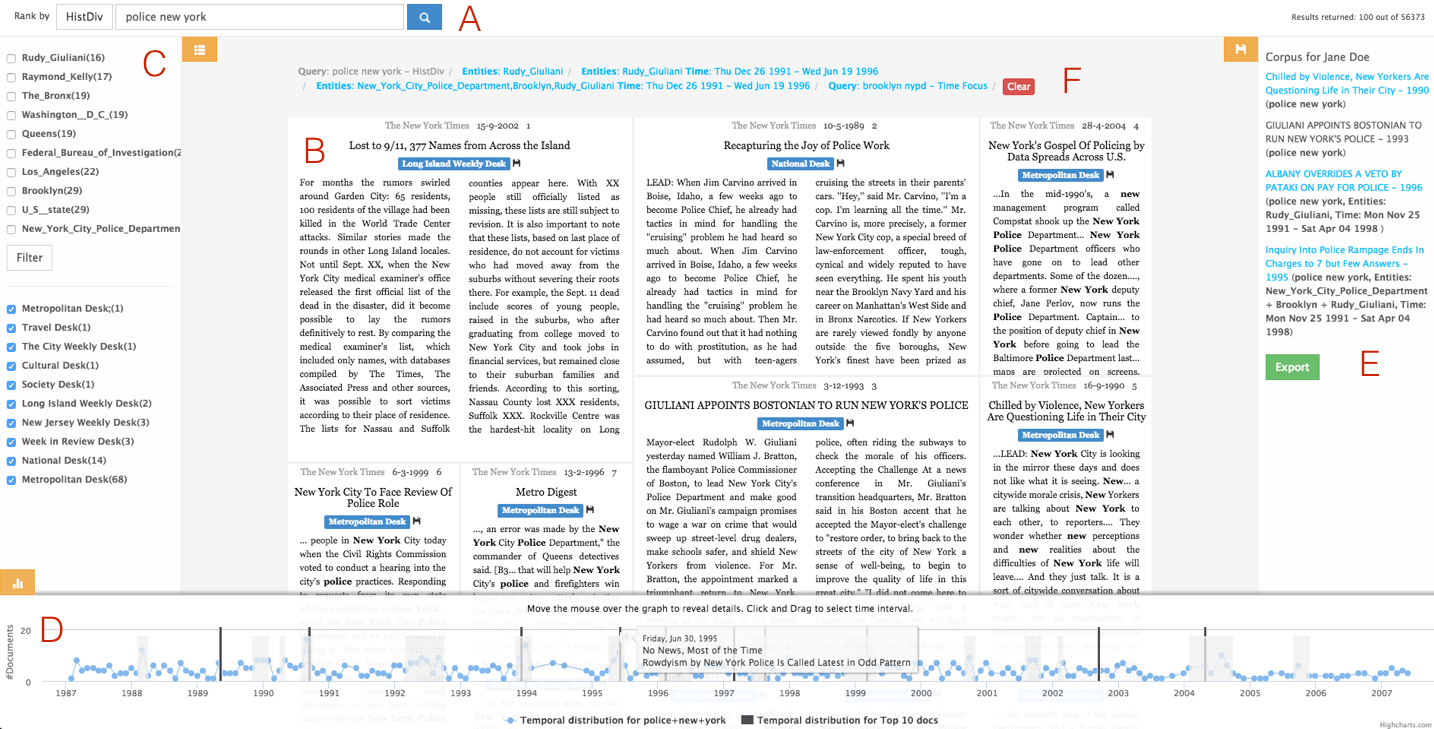} 
    \caption{\textsf{Expedition}'s User Interface: (A) Search box and retrieval model selector (B) Search result list (C) Left pane consisting of entity filter and article type filter (D) Timeline - shaded areas represent bursts (E) Corpus creation pane displaying the articles saved by the user (F) Search trail - tracks the user's actions during corpus creation (the gray text in the trail represents the current state of the user).}
    \label{fig:ui}
\end{figure*}

Consider the scenario of a historian interested in the history of law enforcement in New York City. She has access to the New York Times news archive via a time-aware exploratory search system. Studies have shown that scholars often use general terms, names of people and places as search queries~\cite{buchanan2005information}. Lets say the user enters the terms `police new york' as the initial query resulting in over 30,000 documents. It is infeasible for her to read all documents to get an overview. Using a diverse ranked list for such a query on the other hand can enable her to get an initial anecdotal overview of the query topic by reading just the top results. In doing so, she finds that Rudolph Giuliani was responsible for reforming the police in the 90s and filters her results to only include articles about Giuliani. 
While exploring the archive, it is also important to identify interesting time periods. Typically timelines are used to visualize periods of high and low activity in relation to the query. While even a simplistic timeline is useful for exploration, a more insightful visualization would also include annotations that explicitly highlight important periods and explain why they are important. Based on these annotations, she makes a more educated decision to further investigate the year 1994 - the year Giuliani came to power and vowed to reform the police. In this new list of documents she encounters several interesting articles which she would like to further analyze offline. She uses a simple save button to assign articles to an arbitrary corpus. The system also keeps track of which queries, retrieval models and filters she used when building the corpus. Using this documentation, she can easily share her corpus construction procedure and also revisit certain stages to explore in different directions. 

To support such a scenario an ideal system should offer (i) a host of retrieval models, both diversity aware and ad-hoc, to support varying intents (ii) appropriate filters to refine the intent (iii) an insightful visual overview across time (iv) the ability to construct corpora and (v) the ability to document the exploration process. In this work we describe \textsf{Expedition} (Figure~\ref{fig:ui}), a time-aware exploratory search system built for scholarly exploration in archives. \textsf{Expedition} allows users to enter keyword queries and vary the intent by choosing an appropriate retrieval model. The results are displayed using an adaptive layout that gives more screen real estate to more relevant results. To aid exploration in time, the user is presented with a timeline which is annotated to indicate and describe important time periods of the query. Users are also presented with a list of salient entities to the query which serve as filters. Every query and subsequent refinement is stored and displayed in the search trail - documenting the user's progress. Users can also navigate back and forth through their exploration process by clicking on a particular state. Finally, users can add documents to their corpus and export it at any point.

\section{System Design}


The \textsf{Expedition} UI is shown in Figure~\ref{fig:ui}. At the top of the interface the user can enter her keyword query and further specify her intent by selecting an appropriate retrieval model from a drop down list consisting of textual relevance, temporal relevance, topical diversity, temporal diversity and historical diversity. The search results are displayed using an adaptive newspaper layout where article real estate is directly proportional to its' rank. The timeline which is found at the bottom is generated using a combination of publication dates and temporal references from the pseudo-relevant documents. This timeline is further annotated with bursts which are labeled to provide the user with more insights. In the same timeline we also visualize the position of the top-k documents, allowing the user to examine the temporal scope of the top results. Furthermore, a list of salient entities can be found to the left of the results which can be used as filters. Every time the user re-specifies her intent, her actions are added to the search trail to help keep track of exploration process. Finally, the user can easily add documents to her corpus by clicking on the save button found below the article headline. \textsf{Expedition} can be accessed online at the following url: \texttt{http://bit.ly/archive-search}

In the remainder of this section we detail our system architecture and design choices made when creating \textsf{Expedition}.

\subsection{Data Preparation}

For the purpose of this prototype we used the New York Times Annotated Corpus which is a news archive consisting of 1.8M articles spread across 20 years. To support sophisticated retrieval models and generate insightful timelines, we first semantically enrich the corpus. Though the corpus already contains manual annotations for some entities, we additionally disambiguated all entities in the corpus using a state-of-the-art named entity disambiguation system~\cite{yosef2011aida} to improve coverage. We also extracted temporal expressions from all articles for timeline construction. The disambiguated entities are also useful for helping the user specify her intent during query time. For instance if the user is querying for \texttt{george bush}, she can specify whether she meant George Bush Sr. or George W. Bush to increase the accuracy of the result list. We indexed the entities, publication dates, article content from the augmented collection using Apache Lucene~\footnote{http://lucene.apache.org}.


\subsection{Architecture}

We created a REST API with different end points to support the three major components of the system: retrieval and ranking, timeline generation and salient entity selection. This allows for the decoupling of the user interface which is simply a REST client that interacts with the API and visualizes data. The state of exploration displayed in the search trail and the corpus is maintained on the client side. In the remaining subsections we detail the algorithms used for each component. 

\subsection{Retrieval and Ranking}

\textsf{Expedition} is equipped with three classes of retrieval models: textual relevance, temporal relevance and diversity. While textual and temporal relevance help users focus their results, the diversification algorithms are better suited to overviews and multifaceted queries. Depending on the intent, the user can use a combination of keywords and retrieval model to get the better results. \cite{mishra_expose:_2015} is another system that offers users a suite of temporal retrieval models to choose from. However, the retrieval models are designed for inputs consisting of both keywords and time expressions which is better suited to event exploration. To this end we include the following retrieval models:

\textbf{Temporal Relevance}: Textual relevance does not ensure that the user gets relevant results from the most important time periods. For instance, when searching for \texttt{world trade center}, results from 2001 are perceived to be more relevant than other years. We use the temporal language model suggested in~\cite{li2003time} to help users focus their results to more important time periods.
\\
\textbf{Temporal Diversity}: The algorithm suggested by~\cite{berberich2013temporal} extends the temporal language model by adding an exponential decay to greedily select textually relevant documents from previously uncovered time periods. This retrieval model is particularly useful in getting an overview of a query across time. When searching for \texttt{world trade center}, results from both 2001 and the 1993 bombing are returned using this retrieval model. 
\\
\textbf{Topical Diversity}: Topical diversification is especially useful when the user wants a topical overview of the query. Topical diversity is generally used for ambiguous intents but it can also be used for queries with a multi-faceted information need. For example, when searching for \texttt{David Beckham}, the results should cover his career at Manchester United and Real Madrid. Topical diversity however is time agnostic which increases the likelihood of not getting a sufficient overview across time. We use Ia-Select~\cite{agrawal2009diversifying} as our topical diversification algorithm. Ia-Select relies on modeling the content of a document using aspects. In our system we use entities as aspects.
\\
\textbf{Historical Diversity}: To get a historical overview for a query we use HistDiv~\cite{histDiv} - a diversification approach that jointly optimizes coverage of the most important aspects from the most important time periods. When searching for \texttt{David Beckham}, the top results include articles about his best achievements at Manchester United and Real Madrid when he was playing for them rather than articles that retrospectively looked back at his career. Such an overview serves as an ideal starting point to spur further exploration. 


\subsection{Timeline}

For a more visual overview and to identify important time periods, we incorporate a timeline based visualization in \textsf{Expedition}. A timeline is an essential feature in all time-aware exploration systems. Timelines have been generated by plotting the number of documents in a result set aggregated by the publication date~\cite{matthews2010searching}, temporal expressions found~\cite{alonso_clustering_2009} or both. In \cite{Alonso:2010:NNE:2128344.2128429}, Alonso et al. devised a system with two timelines: a coarse global event timeline and a fine grained timeline plotting the top results. Matthews et al. \cite{matthews2010searching} use a single timeline to visualize document changes across time as well as the change in entity mentions over time. A sliding window across this timeline can be used to zoom into a period. A second timeline is then used to visualize the headlines from top articles published in a selected period. \cite{setty2010inzeit} also uses a timeline to explore the collection, however they visualize only insightful time points rather than plotting a temporal profile of the query. 

Estimating an insightful timeline is an important challenge. In \textsf{Expedition}, we use a combination of temporal references and publication dates to compute a more accurate timeline. While certain time periods experience high publication activity, some other periods maybe highly referred to and are also interesting to scholars. To generate a timeline, we first generate a probability distribution across time by discretizing time and aggregating the documents published in each time unit. We then smooth this distribution using the probability distribution generated by temporal references. We consider the top 1000 documents ranked by the unigram language model as pseudo relevant and utilize a monthly granularity for aggregations. To detect insightful periods or bursts we use the burst detection algorithm suggested in~\cite{peetz2014using}. A burst is highlighted by a shaded region over the timeline. On mouse-over, the label of the burst is displayed. The burst label consists of the headlines of the top three articles published during the burst. This way users can make a more educated decision when selecting which period to focus on. Additionally, we plot the position in time of the top 10 articles. This gives the users an idea of the time periods that the top results cover and whether interesting periods have been represented in the top results or not.

\subsection{Time and Entity Filters}

We forsake standard filtering functionality for time and entity \emph{selectors}. Applying a time period or entity filter causes a new query to be issued with the updated constraints. The result is a list of articles with filtered results from the previous search at the top followed by more articles matching these constraints. Consequently, a new timeline is generated and new entities are added to the left pane to spur further exploration. 

\textbf{Time Selector}: The timeline doubles as an interactive time period selector. Users can select a time period by clicking on a burst directly or by clicking and dragging across the timeline. In its current state, \textsf{Expedition} supports document retrieval and ranking from a specific time period only based on publication dates which are time points rather than intervals. This scenario does not require temporal indexing techniques such as~\cite{Anand:2011:TIS:2009916.2009991}, which are more effective when utilizing temporal expressions, for faster retrieval. In principle, we could employ temporal indexing techniques for collections with documents with valid time intervals.
\\
\textbf{Entity Selector}: We include a list of salient entities to the left of the result list. The entities give the user an overview from a different perspective and also allow the user to refine her intent. To compute the set of top-k salient entities (10 in our prototype) for a given query, we use the following steps: first we select the top 100 documents returned by the chosen retrieval model; we then compute the salient entities for each article using the method suggested in~\cite{fetahu2015automated} and sort by document frequency; finally, we select the top 10 frequent entities from the sorted list. 


\subsection{Search Result Display}

Typically scholars require more context compared to standard snippets when deciding which articles to add to their corpus. News Archive exploration systems like \cite{matthews2010searching,setty2010inzeit} display the result headline and fixed-length snippet in a traditional list view. We, on the other hand, use the newspaper layout suggested in ~\cite{golovchinsky1997newspaper} - \emph{a newspaper represents a mature information medium that is well suited to relatively short and loosely related pieces of text}. The newspaper style interface is well suited to exploring large archived collections since it supports: a multi-narrative scenario, comparison of articles side-by-side and varying degrees of importance by varying the article size. This type of result visualizations is also useful when displaying diversified results. In \textsf{Expedition}, we alter the size of the article depending on its' rank based on the assumption that more relevant results need larger previews. We compute these text previews using Lucene's built-in snippet generator. Clicking on the headline of any result brings up the full article. At the end of each article we add an entity selector consisting of the salient entities mentioned in the document. Lastly, each article also has a save button (floppy disk icon) so that scholars can create a corpus of study.

\subsection{Search Trail and Corpus Building}

The search trail is a series of consecutive stages that reflect the user's exploration. A new stage is appended to the trail each time the user selects a time period or entity, changes the retrieval model or alters the query. Each stage can be revisited by clicking on the corresponding entry in the search trail. 
To the right of the search result list is the corpus creation panel. When the user begins a new session, a new corpus is created automatically. The purpose of this panel is to display all documents the user saved. We display the headline of each saved article along with the query used to retrieve it. Clicking on a headline redirects the user to the full article. We also add an export button to this panel to allow users to download their corpora and search trail for offline analysis.

\section{Conclusion \& Future Work}

In this paper we have described the design and implementation of \textsf{Expedition} - a time-aware exploratory search systems for scholars rather than general users. \textsf{Expedition} provides scholars with a multitude of retrieval models, an insightful timeline visualization, salient entity selectors and corpus creation facilities in a unified system. In the future, we plan to add query suggestions based on previously seen articles. We will also add support for multiple queries to be visualized in the same timeline to make temporal comparisons easier. At present \textsf{Expedition} supports only the manual selection of retrieval model. An interesting direction for further research is the automatic suggestion of retrieval models depending on the state of the user in the session. \textsf{Expedition} can also be used for a variety of different archives, not just news. Web archives are steadily gaining popularity among scholars yet access methods are lacking. Web archives also pose different problems when compared to news archives which are clean and consistent. We envisage that a version of \textsf{Expedition} designed to support various types of archives could benefit scholars immensely. With the amount of information being gathered in digital and web archives increasing rapidly, systems like \textsf{Expedition} can be commercialized for national libraries, archives and publishing houses. This work was carried out in the context of the ERC Grant (339233) ALEXANDRIA. 


%
\bibliographystyle{abbrv}
\bibliography{sigproc}  
%
%

\end{document}